\documentclass[superscriptaddress,aps,9pt,twocolumn]{revtex4-1}
\usepackage[utf8]{inputenc}
\usepackage{graphicx}
\usepackage{verbatim}
\usepackage{amsmath}
\usepackage{amssymb}
\usepackage{bm}
\usepackage{braket}

\begin{document}

\title{ArQTiC: A full-stack software package for simulating materials on quantum computers}

\author{Lindsay Bassman}
\affiliation{Lawrence Berkeley National Lab, Berkeley, CA 94720}
\author{Connor Powers}
\affiliation{University of Southern California, Los Angeles, CA 90089}
\author{Wibe A. de Jong}
\affiliation{Lawrence Berkeley National Lab, Berkeley, CA 94720}

\begin{abstract}
ArQTiC is an open-source, full-stack software package built for the simulations of materials on quantum computers.  It currently can simulate materials that can be modeled by any Hamiltonian derived from a generic, one-dimensional, time-dependent Heisenberg Hamiltonain.  ArQTiC includes modules for generating quantum programs for real- and imaginary-time evolution, quantum circuit optimization, connection to various quantum backends via the cloud, and post-processing of quantum results.  By enabling users to seamlessly perform and analyze materials simulations on quantum computers by simply providing a minimal input text file, ArQTiC opens this field to a broader community of scientists from a wider range of scientific domains.
\end{abstract}

\maketitle

\section{Introduction}
Quantum computers are an emerging technology, which are poised to revolutionize the computational sciences \cite{alexeev2021quantum, outeiral2021prospects, cao2019quantum, mcardle2020quantum}.  Using quantum bits, or qubits, as the units of information processing, quantum computers can capitalize on purely quantum phenomenon such as superposition and entanglement to achieve exponential speed-ups and memory reductions compared to their classical counterparts for some applications.  Originally conceived of for the simulation of quantum systems \cite{feynman1982simulating}, quantum computers were later rigorously proven to offer a computational advantage in this area \cite{lloyd1996universal, abrams1997simulations, zalka1998simulating}.  Indeed, the simulation of quantum materials is seen as one the most promising applications for quantum computers in the near term \cite{bassman2021simulating}.  Quantum materials are materials in which quantum effects at the microscopic level give rise to exotic phases or other emergent behaviors at the macroscopic level \cite{keimer2017physics}.  An explosion of research into quantum materials over the last decade suggests that such materials will be crucial for the development of next-generation technologies \cite{tokura2017emergent, basov2017towards, giustino20212021}.  Thus, elucidating the properties and dynamics of quantum materials through simulation is a much anticipated milestone for near-term quantum computers.  

At present, the software available for designing and executing simulations of quantum materials on quantum computers is in a nascent stage, often requiring a great deal of domain knowledge in quantum computation.  At the most fundamental level, a program run on a quantum computer is a sequence of physical operations performed on the qubits (e.g., electromagentic pulses).  However, much like writing code for classical computers in binary, writing code for quantum computers in terms of pulses can be cumbersome and difficult.  To alleviate this burden, layers of abstraction can be sequentially added atop the pulse-level programming layer to facilitate writing quantum programs.

\begin{figure}[h]
    \centering
    \includegraphics[scale=0.6]{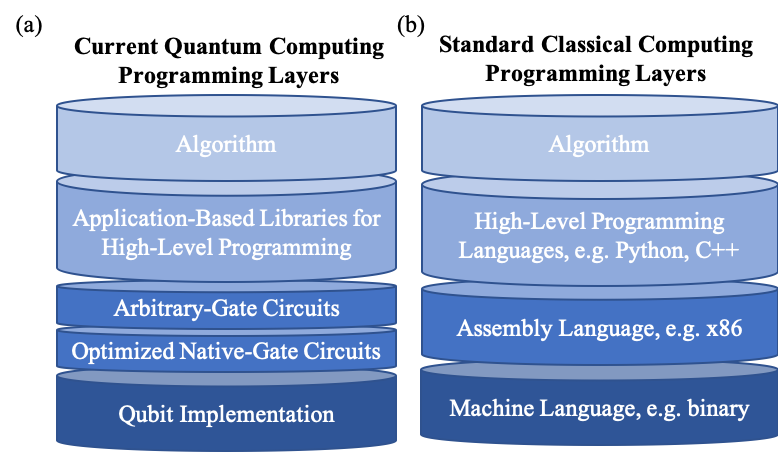}
    \caption{Programming layers for quantum (a) and classical (b) computers.}
    \label{qcSoftwareStack}
\end{figure}

The current hierarchy of sequentially abstracted programming layers for quantum computing is shown in Figure \ref{qcSoftwareStack}a.  At the bottom, the qubit implementation dictates which physical operations can be applied to the qubits.  Abstracting one layer above this involves representing the simulation in terms of an optimized native-gate circuit, which is a serial list of quantum logic gates acting on the qubits, with native gates having a one-to-one correspondence with implementable operations on the qubits.  Sitting a level above native-gate circuits are arbitrary-gate (i.e., any unitary matrix) circuits.  Note that while native-gates are backend-dependent, arbitrary-gates are backend-agnostic, allowing greater flexibility.  Abstracting one layer above gates, simulations can be designed using high-level programming models based on application-focused libraries.  This is currently one of the more under-developed layers.  Finally, at the top resides the algorithm which abstracts away all implementation details and solely describes the general process by which the system of qubits should be manipulated.  Figure \ref{qcSoftwareStack}b shows analogous programming levels in standard classical computers for comparison.  

In the noisy intermediate-scale quantum (NISQ) era \cite{preskill2018quantum}, developing code to run on quantum computers generally involves programming at the gate level.  Furthermore, different quantum backends (either the real quantum processors or quantum simulators, which simulate quantum computers with classical computers) use their own hardware-specific language, making it difficult to port simulation code written for one quantum machine to another.  Together, this presents a large barrier to entry for researchers from other relevant areas of science, such as chemistry, materials science, and condensed matter physics, who can provide a wealth of new perspectives and insights for materials simulations on quantum computers.  To hasten progress, we have developed an open-source, backend-agnostic, high-level programming library called Architecture for Time-dependent Circuits (ArQTiC), to facilitate research in this area for scientists from a diverse range of backgrounds by lowering this barrier to entry.  

As a full-stack software package, ArQTiC provides access to each of the programming layers presented in Figure \ref{qcSoftwareStack}a.  At the top layer, ArQTiC implements two major algorithms for material simulation: (i) Hamiltonian evolution based the Trotter decomposition \cite{lloyd1996universal, trotter1959on}, and (ii) imaginary-time evolution via the quantum imaginary time evolution (QITE) algorithm \cite{motta2020determining}.  The former is useful for studying the dynamic behavior of materials and their properties, while the latter can be used to find ground- and excited-state energies as well as for generating thermal states, which can be used to compute properties of materials at finite temperatures.  

To aid the user in implementing these algorithms, ArQTiC provides a high-level programming library specific to the application of simulating materials on quantum computers.  At the gate-level, ArQTic provides automatic optimization of the quantum circuit using state-of-the-art circuit synthesis and compilation tools, and can translate circuits into several different languages targeting different quantum backends.  Furthermore,  ArQTic can connect with the IBM and Rigetti quantum computers via the cloud to execute the circuits.  Finally, ArQTiC provides post-processing and analysis of the data returned via the cloud from the quantum backend.  The full code, as well as an array of python notebooks demonstrating various simulation uses cases (including the illustrative examples given in Section \ref{sec:examples}), are available on GitHub \cite{github}.  By giving users the ability to easily generate, optimize, execute, and post-process quantum circuits simulating materials on quantum computers, ArQTiC in essence brings this important class of simulations to the masses.  

\section{Software Description}
ArQTiC offers a full-stack solution for the simulation of materials on quantum computers.  Taking as input a simple text file, in which the user defines various simulation parameters, ArQTiC can generate, optimize, and execute circuits for the quantum simulation, as well as post-process experimental results all with only a few high-level functions calls.  This enables researchers from a range of physical sciences to easily perform their own materials simulations on quantum computers without needing to understand the low-level mechanics and intricacies of quantum computation.  

\begin{figure}[h]
    \centering
    \includegraphics[scale=0.3]{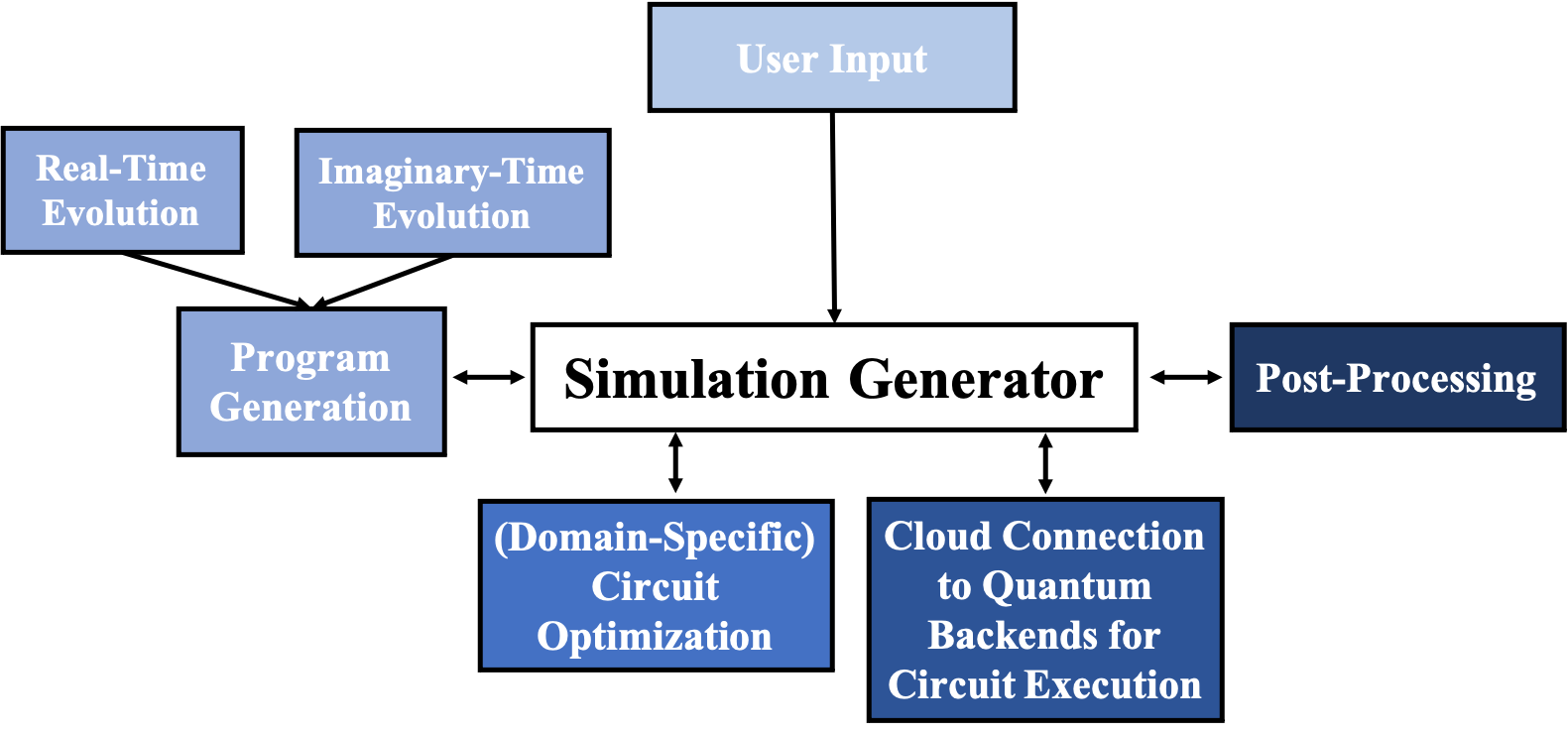}
    \caption{Blueprint diagram ArQTiC modules.  The Simulation Generator is the central data structure, which is instantiated through reading in a user-provided input text file.  The Simulator Generator interacts with all other modules of ArQTiC, including program generation, circuit optimization, connection to quantum backends, and post-processing.}
    \label{fig:blueprint}
\end{figure}

A blueprint of the ArQTiC programming library is shown in Figure \ref{fig:blueprint}, depicting the various modules.  The central data structure is the Simulation Generator, which contains all the information relevant for the simulation.  It is instantiated by reading in a user-provided input text file.  The Simulation Generator is what interfaces with all other modules of ArQTiC to perform a simulation, including modules for program (i.e. quantum circuit) generation, circuit optimization, connection to quantum backends via the cloud, and post-processing.  

\begin{figure*}
    \centering
    \includegraphics[scale=0.7]{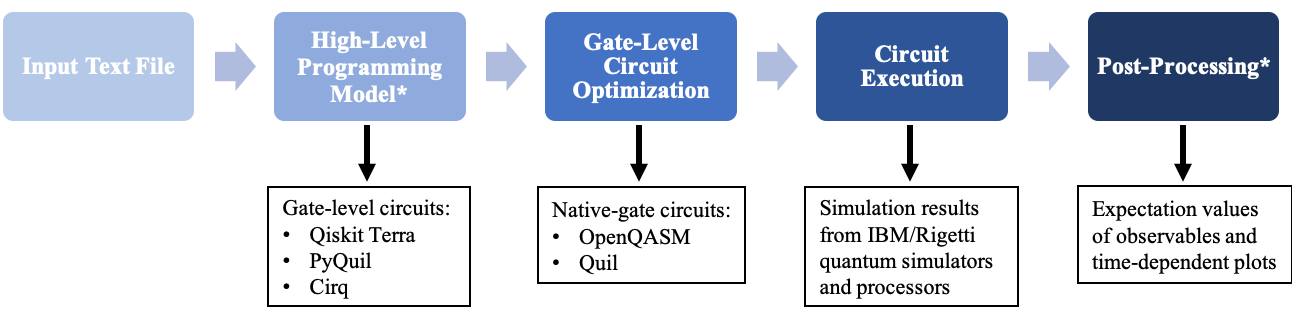}
    \caption{ArQTiC workflow diagram.  The top row of boxes summarizes the flow of information from the input text file, through the high-level programming layer, to the gate programming layer, down to pulse scheduling for circuit execution on the qubits.  The bottom rows shows optional outputs at each stage of the workflow. The starred components denote the main contributions of ArQTiC.}
    \label{fig:workflow}
\end{figure*}

Figure \ref{fig:workflow} shows a workflow diagram, which illustrates how all the modules of ArQTiC come together to seamlessly provide a user with materials simulation results from a quantum backend given just a simple input text file. Note that the boxes in Figures \ref{fig:blueprint} and \ref{fig:workflow} are color coordinated to demonstrate the correspondence between modules in ArQTiC and subsections of the workflow.  Performing a simulation with ArQTiC begins with a simple input text file providing the user-defined parameters of the system Hamiltonian as well as other relevant simulation parameters. Sample input files are provided in the illustrative examples in Section \ref{sec:examples}.  Currently, ArQTiC can generate simulations for materials which can be modeled with a time-dependent Heisenberg Hamiltonian in one-dimension of the form
\begin{equation} \label{hamiltonian}
    H(t) = \sum_{\alpha}\sum_{i=1}^{n-1} J^{\alpha}_i(t)\sigma_{i}^{\alpha}\sigma_{i+1}^{\alpha} +  \sum_{\alpha}\sum_{i=1}^{n} h^{\alpha}_i(t)\sigma_{i}^{\alpha} 
\end{equation}
where $\alpha \in [x,y,z]$, $n$ is the number of spins in the system, $J^{\alpha}_i$ is the time-dependent strength of the exchange interaction between nearest neighbor spins $i$ and $i+1$ in the $\alpha$-direction, $h^{\alpha}_i$ is the time-dependent strength of the external magnetic field in the $\alpha$-direction acting on spin $i$, and $\sigma_{i}^{\alpha}$ is the $\alpha$-Pauli matrix acting on qubit $i$.  The large amount of freedom in defining the parameters in Equation \ref{hamiltonian} makes this Hamiltonian quite versatile in its ability to model a wide range of systems including ubiquitous models such as the transverse field Ising model (TFIM), the XY model, and the XXZ chain.  The parity of the $J^{\alpha}_i$ parameters can be chosen to simulate ferromagnetic or antiferromagnetic systems, while setting their values to be uniform across all spin pairs versus randomly varied allows one to simulate ordered or disordered systems (such as spin glasses \cite{lidar1997simulating}), respectively.  Finally, specifying a time-dependent function for the external magnetic field amplitude, allows one to simulate laser pulses on material of interest\cite{shin2018phonon, bassman2020towards}.  

Once the system of interest has been determined, the user can set the Hamiltonian parameters appropriately in the input file, along with other relevant details of the simulation.  The Simulation Generator object is then instantiated by reading in the input file.  It can then be used to generate programs, which are ArQTiC's native intermediate representation of the quantum circuit that performs the material simulation.  A program is essentially a backend-agnostic list of arbitrary gates.  The advantage of working with this intermediate representation is that the gate-level circuit can be designed once and simply translated into any other language required by a specific backend.  Currently, ArQTiC supports converting its programs into Qiskit \cite{Qiskit}, PyQuil \cite{smith2016practical}, and Cirq \cite{cirq}.  In particular, this makes it easy to run the same simulation on multiple different quantum backends for comparison.  

The creation of a new program relies on algorithms for real- and imaginary-time evolution under a given system Hamiltonian.  Separate modules exist for generating a high-level programs for either real-time evolution based on Trotter decomposition \cite{lloyd1996universal, trotter1959on} or imaginary-time evolution based on QITE \cite{motta2020determining}.  One or the other will be called depending on a boolean flag set in the input file. Real-time evolution can be simulated under a time-independent or time-dependent Hamiltonian with ArQTiC.  Time-independent Hamiltonians are generally used for simulating quantum quenches, whereby the material system is initialized in the ground state of one Hamiltonian, but is made to evolve under a different Hamiltonian \cite{mitra2018quantum, smith2019simulating, sopena2021simulating, fauseweh2021digital}.  Quenching from one Hamiltonian to the other can be viewed as instantaneously changing the environment of the material, thereby altering its Hamiltonian.  These types of simulations aim to answer fundamental questions about  many-body localization, the mechanisms and timescales of thermalization, the changes to or development of collective order (e.g., ferromagentism, superconductivity, topological order) under a quench, the universality of the dynamics in quenches near critical points, and more \cite{mitra2018quantum}. 

Simulations under time-dependent Hamiltonians \cite{poulin2011quantum} can be useful within a few different paradigms.  First, they can be used to simulate dynamic processes, such as scattering \cite{du2020quantum}. Second, they can be used to simulate materials in dynamic environments, such a time-dependent external magnetic field \cite{bassman2020towards, bassman2021computing}.  A third use-case is for finding the ground state of a material through adiabatic quantum evolution \cite{barends2016digitized}.  Here, the material is initialized in the ground state of an initial  Hamiltonian $H_I$, which is presumed to be easy to prepare on the quantum computer.  The material is then evolved under a parameter-dependent Hamiltonian $H(s) = (1-s)H_I + sH_P$, which slowly (adiabatically) transforms from the initial Hamiltonian $H_I$ to the problem Hamiltonian $H_P$ as the parameter $s$ is varied from 0 to 1.  The adiabatic theorem states that if the system is initialized in the ground state of $H_I$ and $s$ is varied from 0 to 1 slowly enough, the system will remain in the instantaneous ground state of the Hamiltonian $H(s)$.  Thus, at the end of the protocol, the system will be in the ground state of the problem Hamiltonian $H_P$, which is, in general, difficult to prepare.  In this way, simulation with a time-dependent Hamiltonian within ArQTiC can be used to generate the ground state and measure the ground-state energy of various Hamiltonians. 

ArQTiC can also generate programs for imaginary-time evolution.  These simulations are useful for two main applications.  The first is for computing the ground-state energy of a material.  As a system is evolved in imaginary time, the lowest energy states begins to dominate the system's wavefunction.  Therefore, simulating the evolution of the material through imaginary time will cause measurements of the system's energy to result in the ground-state energy with higher and higher probability.  The second application for imaginary time evolution is for generating pseudo-thermal states, which can be used to compute properties of materials at finite temperatures.  In particular, these pseudo-thermal states are called minimally entangled typical thermal states (METTS) and are generated in a Markov chain according to the METTS algorithm \cite{white2009minimally}.  On a quantum computer, the QITE algorithm can be used to generate the METTS, via the so-called QMETTS algorithm \cite{motta2020determining}.  The average over measurements of an observable in an ensemble of METTS will give the thermal average of that observable for the material at a given finite temperature \cite{stoudenmire2010minimally}.

Once a program has been created by the Simulator Generator, it must be translated into an optimized, native-gate quantum circuit targeting the quantum backend selected by the user.  In the NISQ era, circuit optimization is equivalent to circuit minimization.  This is because currently available quantum computers suffer from high gate-error rates and short qubit decoherence times, causing simulation results to lose fidelity as the quantum circuit gets larger.  ArQTiC offers several choices for circuit optimization.  The first option uses the native circuit compiler of the chosen target quantum backend.  For example, if the user wishes to run the simulation on the Rigetti quantum computer, ArQTiC will translate the program into a PyQuil circuit and call PyQuil's native compilation function on the circuit.  The second option is to use a popular, state-of-the-art circuit optimizer called tket \cite{sivarajah2020t}.  The final option is a domain-specific option that can produce optimal constant-depth circuits for real-time evolution.  Here, domain-specific refers to the fact that this circuit optimization technique can only be implemented for special system Hamiltonians, which are outlined in Ref. \cite{bassman2021constant}.  For generic Hamiltonians, circuits for real-time evolution are expected to grow at least linearly in size with simulation time.  Due to NISQ-era constraints on circuit size, this in turn limits the length of time that can feasibly be simulated on quantum computers.  The domain-specific constant-depth circuits, however, enable simulations out to arbitrarily long times.

Once an optimized quantum circuit has been generated, the Simulation Generator can connect via the cloud to either the IBM or Rigetti quantum computers, and send the circuits for execution.  Upon job completion, results are sent back via the cloud and stored by the Simulator Generator for post-processing and analysis.  Results from the quantum backend are returned in the form of counts of the number of times each qubit was measured to be 0 or 1.  Thus, post-processing of the data is required to deduce the observable of interest, such as the value of some time-dependent material property.  If requested by the user in the input file, ArQTiC can automatically plot the results and save the figures to file.  

\section{Illustrative Examples}
\label{sec:examples}
\subsection{Dynamic Simulation}
In this example, we demonstrate Anderson localization in a 5-site transverse field XY model when a transverse field is applied randomly across all spins.  The Hamiltonian for the system is given by:
\begin{equation}
    H=\sum_{i=1}^{n-1} (\sigma_i^x\sigma_{i+1}^x + \sigma_i^y\sigma_{i+1}^y) + \sum_{i=1}^{n} b_i \sigma_i^z
    \label{TFXY}
\end{equation}
where $n$ is the number of spins in the chain, $\sigma_i^{\alpha}$ is the $\alpha$-th Pauli matrix acting on spin $i$, and $b_i$ is the strength of the external magnetic field acting on spin $i$ and is randomly selected for each spin from a uniform distribution centered around zero.  The system is initialized with an excitation in the spin-chain, modeled by flipping the first spin to a spin-down while keeping the remaining spins in the spin-up state.  The system is then evolved through time according to Hamiltonian \ref{TFXY}.  To track the displacement of the excitation through time, the excitation displacement \cite{kokcu2021Cartan} is measured at each time-step, given by the observable:

\begin{equation}
    N = \sum_{i=1}^{n} (i-1)\frac{1-\sigma_i^z}{2}
\end{equation}

\begin{figure}[h]
    \centering
    \includegraphics[scale=0.29]{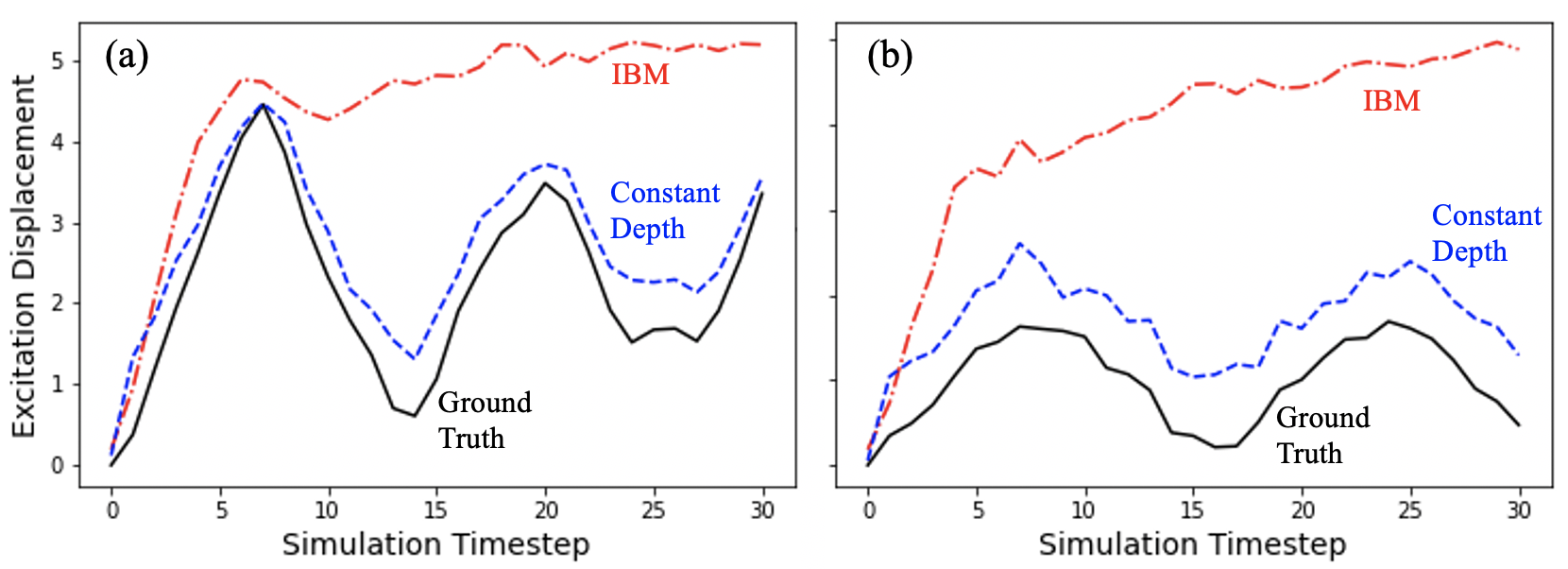}
    \caption{Results of simulating a 5-site transverse field XY spin chain with an initial single-spin excitation with (a) no external magnetic field and (b) randomized transverse external magnetic fields applied to each spin, resulting in localization of the initial excitation. Noiseless ground truth curves are plotted in solid black, results from simulations run on IBM's "ibmq\textunderscore santiago" quantum computer are plotted in dashed red, and results from constant-depth simulations run on the same quantum computer are plotted in dot-dashed blue.}
    \label{fig:dynamics_results}
\end{figure}

Results from dynamic simulations on IBM's ``ibmq\textunderscore santiago" device are shown in Fig. \ref{fig:dynamics_results} for a system with zero external magnetic field (Fig. \ref{fig:dynamics_results}a) and for a system with a randomized external magnetic field drawn from a uniform distribution between -3 and 3 for each spin (Fig. \ref{fig:dynamics_results}b).  As the transverse field XY model is one of the special Hamiltonians in the domain of the constant-depth circuit optimizer, we show simulation results for circuits compiled with IBM's native circuit transpiler (red dot-dashed lines) versus simulation results for the constant-depth circuits compiled with ArQTiC's domain-specific circuit optimizer (blue dashed lines).  For reference, the ground-truth is depicted with the solid black lines.  

As seen in Fig. \ref{fig:dynamics_results}a, when no external magnetic field is applied, the excitation is displaced nearly the length of the spin chain before gradually settling towards the center of the chain. However, when $b_i$ coefficients are pulled randomly from a uniform distribution between -3 and 3, the excitation is confined to oscillating near the beginning of the chain as seen in Fig. \ref{fig:dynamics_results}b, demonstrating the Anderson localization mechanism \cite{kokcu2021Cartan}. 

Fig. \ref{fig:dynamics_results} also showcases the domain-specific, constant-depth circuit optimization capabilities built into ArQTiC based on work presented in Ref. \cite{bassman2021constant}.  Comparing the results from the IBM-compiled circuits to the constant-depth circuits demonstrates the improvement in simulation fidelity achieved with constant-depth circuits.  Importantly, while results from the IBM-compiled quantum circuits do not show significantly different behavior for zero versus random external magnetic fields, the results from the constant-depth circuits do.  Thus, while the constant-depth results may not be exactly quantitatively accurate, they do demonstrate the trend of Anderson localization, while the IBM-compiled results do not.

\begin{figure}[h]
    \centering
    \includegraphics[scale=0.5]{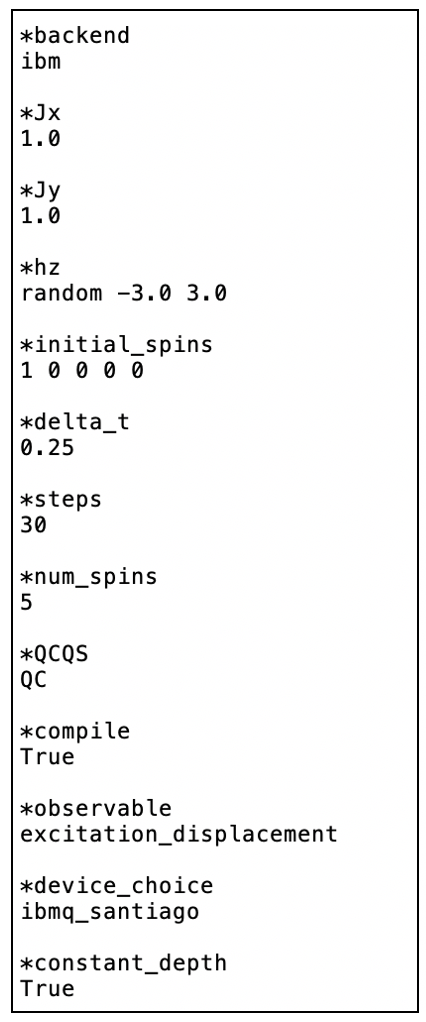}
    \caption{The ArQTiC input file required to perform a noiseless dynamics simulation of excitation displacement within an XY spin chain with a randomized Z-direction magnetic field with strengths between -3 and 3 applied to each spin. The first spin is flipped to create the single-spin initial excitation of this example. This input file would create the ground truth curve shown in Fig. \ref{fig:dynamics_results}b.}
    \label{fig:dynamics_inputfile}
\end{figure}

The input file required to perform a constant-depth dynamics simulation of this system with randomized magnetic field coefficients, visualized by the blue dashed curve in Fig. \ref{fig:dynamics_results}b, is shown in Fig. \ref{fig:dynamics_inputfile}.

The input file shown in Fig. \ref{fig:dynamics_inputfile} can be reconfigured to produce the dot-dashed line in Fig. \ref{fig:dynamics_results}b by either changing the \textbf{constant\textunderscore depth} parameter to ``False" or completely removing the parameter from the input file. If the \textbf{QCQS} parameter is set to ``QS", which will run the simulation on a noise-free quantum simulator, the input file will produce the solid ground truth curve shown in Fig. \ref{fig:dynamics_results}b.  A tutorial for performing this simulation end-to-end can be found on GitHub \cite{github}.

\subsection{QITE Simulation}
In this example, we demonstrate how to use ArQTiC to find the ground state energy of a material via QITE \cite{motta2020determining}.  Our system of interest is a 3-spin TFIM with open boundary conditions.  The Hamiltonian for this system can be written as:
\begin{equation}
    H= J_z\sum_{i=1}^{n-1}\sigma_i^z\sigma_{i+1}^z + h_x\sum_{i=1}^{n}  \sigma_i^x
\end{equation}
where $\sigma_i^{\alpha}$ is the $\alpha$-th Pauli operator acting on spin $i$, $J_z$ gives the strength of the exchange coupling between nearest neighbor spins, $h_x$ gives the strength of the external magnetic field acting uniformly on all the spins, and $n$ gives the number of spins in the system.  The QITE algorithm works by evolving the system through an imaginary time $it \equiv \beta$ by applying the evolution operator $U = e^{-{\beta}H}$.  The difficultly with evolving a system through imaginary time on a quantum computer is that this operator $U$ is not unitary (quantum computers can only perform unitary operators on qubits).  The QITE algorithm is able to generate a unitary approximation to this operator by sequentially building up a quantum circuit with a set of sub-circuits.  Each sub-circuit carries out approximate unitary evolution through an imaginary time-step of size $\Delta \beta$.  The sub-circuit for each subsequent $\Delta \beta$ depends on measured expectation values from the total circuit up to the previous time-step.    

\begin{figure}[h]
    \centering
    \includegraphics[scale=0.5]{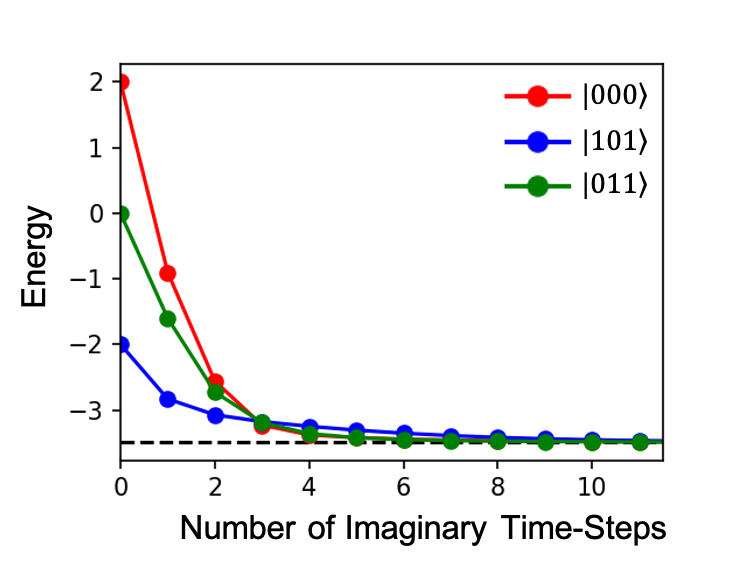}
    \caption{Convergence to the ground state energy (black dashed line) of the 3-spin TFIM spin-chain via QITE.  Different colored lines correspond to different initial product states of the systems.}
    \label{fig:QITE}
\end{figure}

\begin{figure}[h]
    \centering
    \includegraphics[scale=0.5]{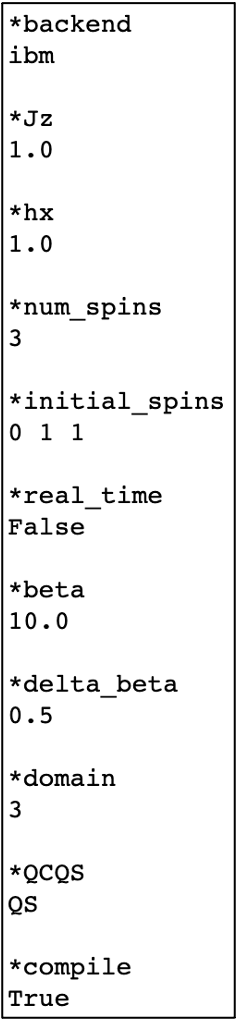}
    \caption{The ArQTiC input file required to perform a ground state energy calculation with QITE. This input file can be used to generate the green curve shown in Fig. \ref{fig:QITE}}.
    \label{fig:qite_code}
\end{figure}

Figure \ref{fig:QITE} shows how the measured final energy of the system (colored, solid lines) converges to its ground state (black, dashed line) as the number of imaginary time-steps is increased.  Different colored lines correspond to starting the system in different initial product states.  As shown, all initial states converge to the expected ground state in about eight imaginary time-steps of size $\Delta \beta = 0.3$.  A sample input file used to generate these results with ArQTiC is shown in Figure \ref{fig:qite_code}.  A tutorial for performing this simulation end-to-end can be found on GitHub \cite{github}.

\section{Conclusion}
We have presented ArQTiC, an open-source, full-stack programming library for performing simulations of materials on quantum computers.  ArQTiC can generate programs (i.e., quantum circuits) for both real- and imaginary-time evolution of a system under a generalized, time-dependent, one-dimensional Heisenberg Hamiltonian.  By constraining certain parameters, this Hamiltonian can be used to simulate various paradigmatic materials Hamiltonians of great interest including the TFIM, the (transverse) XY model, the XXZ chain, and more.  By simply providing the system Hamiltonian parameters and a few other simulation parameters, the user can rely on ArQTiC to seamlessly generate, optimize, and execute materials simulations on various quantum backends, as well as post-process and analyze the quantum results.  The full code, as well as tutorial-style demonstrations of a number of various simulation use cases can be found on GitHub \cite{github}.  By allowing a broader community of scientists to easily perform simulations of materials on quantum computers, ArQTiC paves the way towards accelerated progress in both learning more about quantum materials, as well as designing new quantum algorithms for materials simulations.

\section*{Acknowledgements}
LB and WAdJ were supported by the U.S. Department of Energy (DOE) under Contract No. DE-AC02-05CH11231, through the Office of Advanced Scientific Computing Research Accelerated Research for Quantum Computing and Quantum Algorithms Team Programs. CP was supported as part of the Computational Materials Sciences Program funded by the U.S. Department of Energy, Office of Science, Basic Energy Sciences, under Award Number DE-SC0014607.

\bibliographystyle{apsrev4-1}
\bibliography{main}

\end{document}